\documentclass[hidelinks]{article}

\usepackage{arxiv}

\usepackage{amsmath, amssymb, amsfonts, color}
\usepackage{algorithmicx}
\newcommand{\W}{\mathcal{W}}
\newcommand{\LL}{\mathcal{L}}
\newcommand{\R}{\mathbb{R}}

\newcommand{\norm}[1]{\left\lVert#1\right\rVert}
\newcommand{\abs}[1]{\left\lvert#1\right\rvert}

\newtheorem{thm}{Theorem} 

\newcommand{\ee}{\mathbf{e}}
\newcommand{\xx}{\mathbf{x}}
\newcommand{\yy}{\mathbf{y}}

\newcommand{\bb}{\mathbf{b}}

\newcommand{\uu}{\mathbf{u}}
\newcommand{\UU}{\mathbf{U}}

\makeatletter
\algrenewcommand\ALG@beginalgorithmic{\footnotesize}
\makeatother

\title{Peridynamics for Quasistatic Fracture Modeling}

\author{Debdeep Bhattacharya \\
	Department of Mathematics\\
	Louisiana State University\\
	Baton Rouge, Louisiana 70803\\
    Email: debdeepbh@lsu.edu \\
    and \\
    Patrick Diehl \\    
    LSU Center of Computation \& Technology\\
	Louisiana State University\\
	Baton Rouge, Louisiana, 70803\\
	Email: patrickdiehl@lsu.edu \\
	and \\
    Robert P. Lipton    
    Department of Mathematics \&\\
 LSU Center of Computation \& Technology\\
	Louisiana State University\\
	Baton Rouge, Louisiana, 70803\\
	Email: lipton@lsu.edu
}

\usepackage{siunitx}
\usepackage{graphicx}
\usepackage{xfrac}
\usepackage{subcaption}

\newcommand{\bx}{\mathbf{X}}
\newcommand{\bbR}{\mathbb{R}}
\newcommand{\calH}{H}
\newcommand{\bbK}{\mathbf{K}}
\newcommand{\bA}{\mathbf{A}}
\newcommand{\bE}{\mathbf{E}}
\newcommand{\bu}{\mathbf{U}}
\newcommand{\bzero}{\mathbf{0}}
\newcommand{\be}{\mathbf{e}}

\usepackage{xcolor}
\definecolor{azure}{rgb}{0.0, 0.5, 1.0}
\definecolor{awesome}{rgb}{1.0, 0.13, 0.32}
\definecolor{asparagus}{rgb}{0.53, 0.66, 0.42}
\definecolor{cadetgrey}{rgb}{0.57, 0.64, 0.69}

\usepackage{algorithm}
\usepackage{algpseudocode}

\begin{document}

\maketitle

\begin{abstract}
{\it Fracture involves interaction across large and small length scales. With the application of enough stress or strain to a brittle material, atomistic scale bonds will break, leading to fracture of the macroscopic specimen. From the perspective of mechanics fracture should appear as an emergent phenomena generated by a continuum field theory eliminating the need for a supplemental kinetic relation describing crack growth.
We develop a new fast method for modeling quasi-static fracture  using peridynamics. We apply fixed point theory and model stable crack evolution for hard and soft loading. For soft loading we recover unstable fracture. For hard loading we recover stable crack growth.  We show existence of quasistatic fracture solutions in the neighborhood of stable critical points for appropriately defined energies. The numerical method uses an analytic stiffness matrix for fast numerical implementation. A rigorous mathematical analysis shows that the method converges for load paths associated with soft and hard loading. For soft loading the crack becomes unstable shortly after the stress at the tip of the pre-crack reaches the material strength.}

\end{abstract}



\clearpage
\newpage
\twocolumn

\section{INTRODUCTION}
\label{Introductin}
Peridynamics (PD) is a nonlocal model incorporating force interaction between nearby points within a fixed horizon. In this treatment, the forces are non linear functions of the strain. The force initially increases with increasing strain until a maximum force is reached, and then decreases with increasing strain to zero. Here the strain is formulated as a difference quotient as opposed to a gradient.
This allows the model (the cohesive model) the flexibility to capture fracture as emergent phenomena. It accounts for elastic interaction where the material is intact as well as the emergence and propagation of failure zones. These zones are naturally localized by the model and appear as thin and crack like. This emergent behavior is the hallmark of peridynamic models, see \cite{SILLING2000175,silling2007peridynamic}. It is  theoretically seen for cohesive PD that in the limit of vanishing non-locality the failure zone localizes to surfaces and elastic behavior of intact material surrounding the propagating crack agrees with Linear Elastic Fracture Mechanics (LEFM) \cite{jhalipton2020,liptonjha2021}. It is also seen that the potential energy of cohesive PD converges to the Griffith Energy of LEFM \cite{lipton2014horizonlimit,lipton2016cohesive,lipton2019complex}. Peridynamics was successfully used for the comparison against various experiments\cite{diehl2019review,diehl2021comparative}.

This article addresses theory and numerics of quasi-static fracture using cohesive PD. For the quasi-static case, there is no inertia and time is represented by a load parameter. Our analysis shows (for the first time) an existence theory for quasi-static PD fracture modeling that holds for both hard and soft loading. Here, a quasi-static PD fracture evolution is shown to exist in a neighborhood of a stable critical point of the cohesive PD energy for both hard and soft loading, see  section \ref{Existence}. 
As for numerics it appears that there are only a few quasi static approaches to PD simulation available~\cite{huang2015improved,mikata2012analytical,zaccariotto2015examples,wang2019studies,breitenfeld2014quasi,kilic2010adaptive,rabczuk2017peridynamics,freimanis2017mesh}.
 One significant reason why there are only a few quasi-static PD simulations is the added computational expense in going from dynamics and explicit time integration $\mathcal{O}(n^2)$ to quasi- statics and implicit time integration $\mathcal{O}(n^4)$, where $n$ is the number of discrete PD nodes. The major expense here is the assembly of the tangent stiffness matrix $\mathcal{O}(n^4)$.

Several methods were proposed to speed up the time integration.
Finite element approaches (FEM) for PD~\cite{chen2011continuous,emmrich2007peridynamic,macek2007peridynamics} were applied and found to reduce the computational costs for the assembly of the tangent stiffness matrix to $\mathcal{O}(n^3)$. For purely elastic problems  Wang~\cite{wang2012fast} developed a Galerkin method that exploits the matrix structure and reduces the costs of solving the matrix system from $\mathcal{O}(n^3)$ to $\mathcal{O}(n\log^2(n))$. In another direction, Chen~\cite{chen2011continuous} proposed a simplified model to reduce the computational costs to $\mathcal{O}(n)$ but with a reduced convergence rate of only first order for linear (FEM). Prakash~\cite{prakash2020multi} presents an algorithm using sparse matrices for the assembly of the tangent stiffness matrix instead a dense matrix. The performance of the sparse implementation is compared with an adaptive dynamic relaxation scheme (ADR) in~\cite{kilic2010adaptive}. It is found that a speed-up factor between \num{12} and \num{22} against the ADR solve is achievable. \\

The paper is structured as follows: Section~\ref{Background} introduces the ingredients of cohesive PD model. Section~\ref{Existence} describes the quasi-static crack evolution for soft and hard loading. Section~\ref{Algorithms} introduces the numerical algorithm for the soft and hard loading cases. Section~\ref{Numerical} shows some preliminary simulation results for the hard and soft loading. In ~\ref{Conclusion} we provide conclusions. 

\section{Background}
\label{Background}

We consider quasi-static evolution for the cohesive PD model \cite{lipton2014horizonlimit,lipton2016cohesive,lipton2019complex}. In preparation for the next section that describes the existence of quasi-static evolutions, we introduce the energies associated with peridynamic deformations and Euler Lagrange equations. The deformation field inside the deforming body represented by the domain $D$ is given by $\uu(\xx, t)$.

The strain $S(\yy, \xx, \uu)$ between the point $\xx$ and  $\yy$ is given by
\begin{align*}
    S(\yy, \xx, \uu) = \frac{\uu(\yy) - \uu(\xx)}{\abs{\yy - \xx}} \cdot \ee_{\yy - \xx},
\end{align*}	
where
$
\ee_{\yy - \xx}  = \frac{\yy - \xx}{\abs{\yy - \xx}}
$.
The peridynamic potential energy is
\begin{equation}
    \label{eq:energy}
    PD^\epsilon[\uu] =  \int\limits_{D}^{}  \int\limits_{H_\epsilon(\xx) \cap D} \abs{\yy - \xx} \W^\epsilon(S(\yy, \xx, \uu)) d\yy d\xx.
\end{equation}
Here, $H_\epsilon(\xx)$ is a peridynamic neighborhood of radius (also referred to as the peridynamic \textit{horizon}) $\epsilon$ centered at $\xx$. Here the energy of interaction is given by
\begin{equation}
    \label{eq:energydensity}
    \W(S(\yy, \xx, \uu)) = \frac{J^\epsilon(\abs{\yy - \xx})}{\epsilon^{d+1}w_d \abs{\yy - \xx}}  h(\abs{\yy - \xx} S^2(\yy, \xx, \uu))
\end{equation}	
Where, $J^\epsilon(r) = J(\frac{r}{\epsilon})$, and $J$ is a non-negative bounded function supported on $[0,1]$. $J$ is also called the \textit{influence function} as it determines the influence of the bond force of peridynamic neighbors on the center of the peridynamic horizon as a function of distance. The volume of a unit ball is $w_d$, where $d$ is the dimension $d=2$ or $3$, and  $h$ is a concave function that is thrice differentiable.

\begin{figure}[tb]
     \centering
     \includegraphics[width=\linewidth]{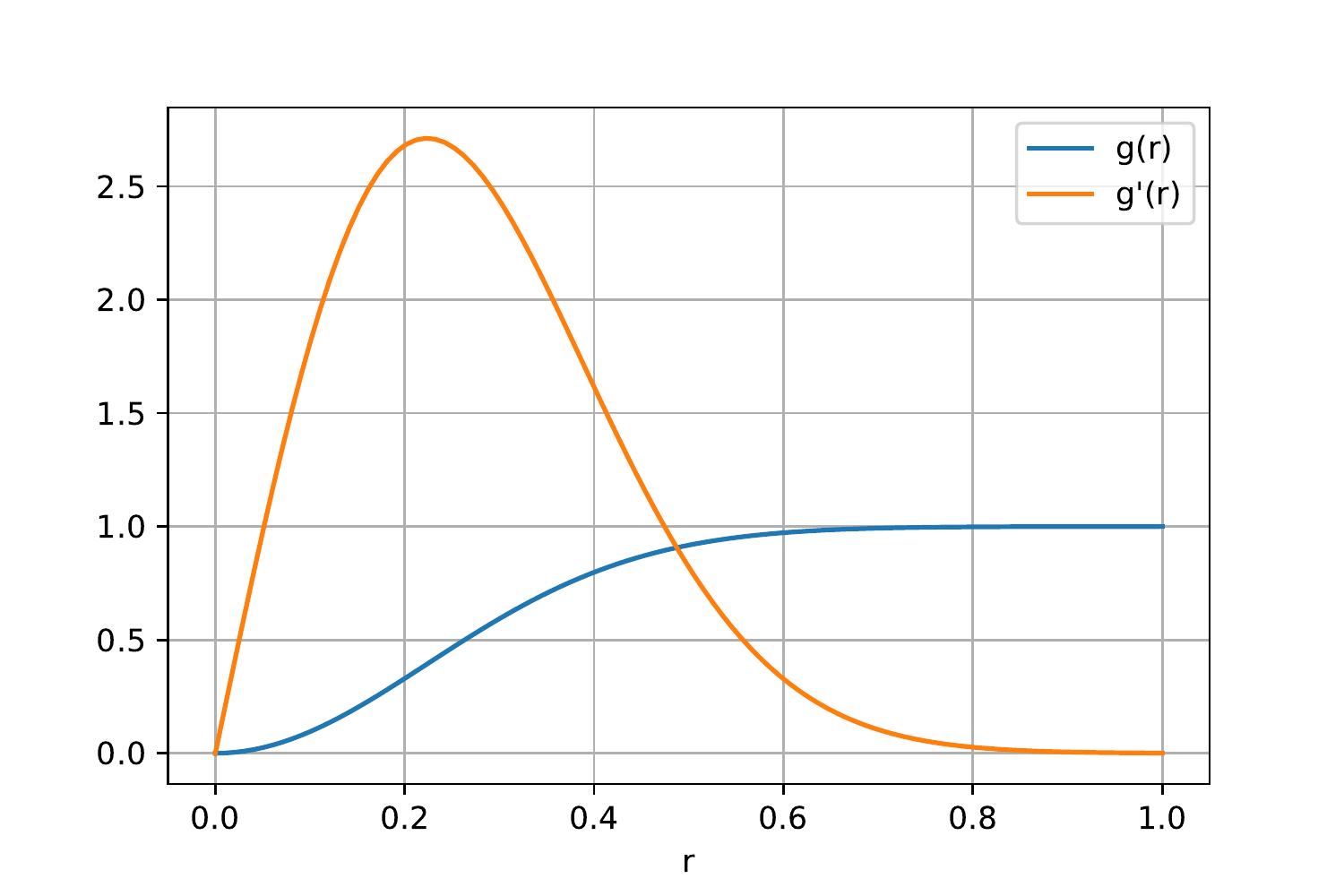}
     \caption{Sketch of the potential $g(r)$ and its derivative $g'(r)$. Before its maximum $g'(r)$ stays in the linear regime and softens after to zero.}
     \label{fig:sketch}
 \end{figure}

The total energy of the system given by
\begin{equation}
    \label{eq:total_energy}
    E[\uu] = - PD^\epsilon[\uu] + \int\limits_{D}^{} \uu \cdot \bb \ d\xx.
\end{equation}	
The critical point of the total energy $\uu^\epsilon$ satisfies the Euler Lagrange
 equation
\begin{equation}
    \label{eq:first-var}
    \nabla PD^\epsilon[\uu^\epsilon] = \bb
\end{equation}	
in alternate notation this is written
\begin{equation}
    \label{eq:equlib}
    \LL[\uu^\epsilon] = \bb,
\end{equation}
where 

\begin{align}
    \LL[\uu](\xx) = &-\int\limits_{H_\epsilon(\xx) \cap D}^{} \frac{2 J^\epsilon(\abs{\yy - \xx} )}{\epsilon^{d+1} w_d \sqrt{\yy - \xx}} \notag \\
    & g'(\sqrt{ \abs{\yy - \xx}} S(\yy, \xx, \uu)) \ee_{\yy - \xx} d\yy,
\end{align}	
with $g(r)=h(r^2)$ or $g$ given by a cubic spline interpolation with prescribed slope at one knot at origin and another knot at the horizontal asymptote. A common example of double well potential $g$ is:
 \begin{equation}
     g(r) = C(1 - \exp[-\beta r])
    \label{eq:potential}
 \end{equation}
 where $C, \beta$ are material dependent parameters. For the energy equivalence to classical theory we refer to~\cite{diehl2016numerical}. Figure~\ref{fig:sketch} sketches the potential $g(r)$ and its derivative $g'(r)$. Before its maximum $g'(r)$ stays in the linear regime and softens after to zero.
 
Next, we prescribe boundary values of a deformation for the non-local model.
Denote the $\epsilon$-extension of the peridynamic domain $D$ by $\overline{D}_\epsilon$ defined as
\begin{equation}
    \overline{D}_\epsilon = \left\{ \xx \in \R^3 : \abs{\yy - \xx} \le \epsilon \text{ for some } \yy \in D \right\}.
\end{equation}	
The peridynamic boundary $\partial D_\epsilon$ of $D$ is defined as
\begin{equation}
\partial D_\epsilon = \overline{D}_\epsilon \setminus D.
\end{equation}	

The peridynamic energy adapted to the assignation of peridynamic boundary values is written
\begin{equation}
    \label{eq:energy-bdry}
    \overline{PD}^\epsilon[\uu] =  \int\limits_{D}^{}  \int\limits_{H_\epsilon(\xx)} \abs{\yy - \xx} \W^\epsilon(S(\yy, \xx, \uu)) d\yy d\xx.
\end{equation}
Note we extended the inner integrand so that $\yy$ can range over $\overline{D}^\epsilon$.

A critical point of the peridynamic energy $\uu^\epsilon$ for deformations $\uu$ such that $\uu = \UU$ on $\partial D_\epsilon$ satisfies the Euler Lagrange
 equation
\begin{equation}
    \label{eq:first-var-bdry}
    \nabla \overline{PD}^\epsilon[\uu^\epsilon] = 0,
\end{equation}	
with $\uu^\epsilon =\UU$ on $\partial D_\epsilon$.
This is written as
    \begin{equation}
    \begin{aligned}
    \begin{cases}
	\LL[\uu^\epsilon]  = 0 &\text{ on } D \\
	\uu = \UU & \text{ on } \partial D_\epsilon.
    \end{cases}
    \end{aligned}	
    \end{equation}

\subsection{Discretization}
To discretize the peridynamic equations of motion, a finite difference approximation is utilized. A set of mesh nodes $\{ \bx_i\in \bbR^d\}_{i=1}^N\subset D$ is placed in the domain $D$. To each of the nodes a surrounding volume $\{V_i\in\mathbb{R}\}_{i=1}^N$ is associated. These volumes are non-overlapping $V_i \cap V_j = \emptyset$ and recover $\sum_{i=1}^N V_i \approx |D|$ the volume of the domain $D$. Each node $\bx_i$ interacts with all neighbors within the finite neighborhood $\calH_\epsilon(\bx_i)=\{\bx_j \, \vert \, \vert \bx_j - \bx_i \vert \leq \epsilon \}$. 

\subsubsection{Description of the analytic stiffness matrix}
The analytic description of the discrete tangent stiffness matrix reads as
\begin{align}
\mathbb{K}(\mathbf{u}) &= \begin{bmatrix}
    \mathbb{K}_{11} & \mathbb{K}_{12} & \ldots & \mathbb{K}_{1N-1} & \mathbb{K}_{1N}\\
    \mathbb{K}_{21} & \mathbb{K}_{22} & \dots & \mathbb{K}_{2N-1} & \mathbb{K}_{2N} \\
    \vdots &  \vdots & \ldots  & \vdots & \vdots  \\
    \vdots & \vdots  & \ldots  &  \vdots & \vdots\\
    \vdots & \vdots  & \ldots  & \vdots &  \vdots  \\
    \mathbb{K}_{N-11} & \mathbb{K}_{N2} & ... & \mathbb{K}_{N-1N-1} & \mathbb{K}_{NN-1} \\
    \mathbb{K}_{N1} & \mathbb{K}_{N2} & ... & \mathbb{K}_{N-1N} & \mathbb{K}_{NN} \\
\end{bmatrix},
\end{align}
where each entry $\bbK_{ij}$ is a $d\times d$ matrix where $d=1,2,3$ is the dimension of the problem. 
We have
\begin{align}
    \bbK_{ij} &= \begin{cases}
    \bA_{ij}, \quad \text{if } i\neq j,\\
    \sum_{\substack{\bx_k \in \calH_\epsilon(\bx_k), \\
    \bx_j \neq \bx_i}} \bA_{ik}, \quad \text{if } i = j,
    \end{cases}
    \label{eq:matrix:cases}
\end{align}
where second order tensor $\bA_{ij}$ is given by
\begin{align}\label{eq:tensorA}
    \bA_{ij} &\equiv \bA_{ij}(U^{k-1}) = \frac{2}{\epsilon^{d+1}\omega_d} \frac{J^\epsilon(|\bx_j-\bx_i|) }{|\bx_j - \bx_i|} \notag \\ 
    &g''(\sqrt{|\bx_j - \bx_i|} S(\bx_j, \bx_i; \bu^{k-1})) \bE_{\bx_j - \bx_i} V_j,
\end{align}
when $i \neq j$, and $\bA_{ii} = \bzero$. Here by the notation $S(\bx_j, \bx_i; \bu^{k-1})$ we mean
\begin{align}
    S(\bx_j, \bx_i; \bu^{k-1}) = \frac{U^{k-1}_j - U^{k-1}_i}{|\bx_j - \bx_i|} \cdot \frac{\bx_j - \bx_i}{\abs{\bx_j - \bx_i}},
\end{align}
for discrete problems, where $\bE_{\bx_j - \bx_i} = \be_{\bx_j - \bx_i} \otimes \be_{\bx_j - \bx_i}$.



\section{Existence of quasi-static evolution about stable critical points}
\label{Existence}

The existence theory is given in the set of bounded displacements $\uu(\xx,t)$, i.e. there exists an interval $0\leq t< T$ and a  bound $R<\infty$ such that
\begin{equation}
    \norm{\uu(t)} = \sup_{\xx \in D} \abs{\uu(\xx,t)}<R,
\end{equation}	
for all $0<t<T$.

Soft loading is defined to be the application of body force  $\bb$  in the absence of prescribed boundary data. We say that the displacement $\uu(t)$ satisfies the quasi-static evolution problem for soft loading with prescribed load path $\bb(t)$, $0\leq t<T$ if it is bounded and satisfies
\begin{equation}
    \label{eq:equlib2}
    \LL[\uu(t)] = \bb(t),
\end{equation}
for $0\leq t<T$.

\begin{thm}[Local existence for soft loading]
    \label{thm:exitence-of-perturbation}
    Let the bounded displacement $\uu_0$  be a critical point of the total energy $E[\uu]$ given by \eqref{eq:total_energy} for the choice $\bb = \bb_0$,
    and if $\uu_0$ is a stable critical point for the total energy $E[\uu]$, i.e.,
    \begin{eqnarray}
	\nabla E[\uu_0] = 0, \\
	\nabla^2 E[\uu_0] > 0.
    \end{eqnarray}	
    Then, there exists an $R>0$ and $T>0$ such that for any load path $\bb(t)$, $0\leq t
     <T$ starting at $\bb_0$ for $t=0$ and $\norm{\bb(t)-\bb(0)}<R$ one has a bounded solution $\uu(t)$
    of
    \begin{equation}
    \label{eq:equlib-t}
    \LL[\uu(t)] = \bb(t),
\end{equation}
for $0\leq t<T$.
\end{thm}

On the other hand, hard loading occurs when the displacement is prescribed on the boundary of the body in the absence of body force.
An evolution $\uu(t)$ for a prescribed boundary displacement load path $\UU(t)$, $0\leq t<T$ exists if it is bounded and satisfies
    \begin{equation}
    \begin{cases}
	\LL[\uu(t)] = 0 &\text{ on } D \\
	\uu(t) = \UU(t) & \text{ on } \partial D_\epsilon 
    \end{cases}
    \end{equation}	
    with $\uu(0) = \UU_0$. We state the local existence of quasi-static solution for hard loading.    

\begin{thm}[Local existence for hard loading]
    \label{eq:conti-hard}
    Let $\UU_0$ be a bounded function supported on $\partial D_\epsilon$. Let ${\uu_0}$ have $\UU_0$ as boundary data and be a stable critical point for the peridynamic potential energy $PD^\epsilon[\uu]$, i.e.
    \begin{eqnarray}
	\nabla \overline{PD}^\epsilon[{\uu_0}]  = 0 \\
	\nabla^2 \overline{PD}^\epsilon[{\uu_0}]  > 0.
    \end{eqnarray}	
    Then, there exists $R > 0$ and $T>0$ such that for all hard load paths $\UU(t)$  in $\Vert\UU(t)-\UU_0\Vert<R$ for $0\leq t<T$ with $\UU(0)=\UU_0$ there exists a unique solution path $\uu(t)$ in  $\Vert \uu(t)-\uu_0\Vert_\infty< R$ such that for all $0\leq t< T$
    \begin{align*}
    \begin{cases}
	\LL[\uu(t)] = 0 &\text{ on } D \\
	\uu(t) = \UU(t) & \text{ on } \partial D_\epsilon 
    \end{cases}
    \end{align*}	
    with $\uu(0) = \UU_0$.
\end{thm}


We point out that both the existence of hard and soft loading are proved using fixed point methods. These methods also prove convergence of the numerical algorithm used here.

\section{Numerical algorithms}
\label{Algorithms}
Algorithms ~\ref{algo:bondbased:solver} and ~\ref{algo:newton:hard} outline the steps to solve for the displacement using a Newton method for hard and soft loading.

\begin{algorithm}[h!]
\begin{algorithmic}[1]
\State Define the external force density $\bb$ and tolerance $\delta$
\State Guess the initial displacement $\mathbf{u}_0$
\State Compute residual $r= \Vert \mathbf{F} \Vert$ with $\mathbf{F}=-\bb(\bx_i) - \mathcal{L}(U^{k-1})(\bx_i)$ 
\While{$r \geq \delta$ }  
\State Assemble the tangent stiffness matrix $\mathbb{K}(\mathbf{u})\in\mathbb{R}^{d\cdot N\times d\cdot N}$ 
\State Remove all columns/rows in $\mathbb{K}$ and $\mathbf{F}$ for nodes with prescribed displacement \label{alg:remove:u}
\State Solve the reduced system $\mathbb{K} \Delta \mathbf{u} = \mathbf{F}$ 
\State $\bu += \Delta u$
\State Compute residual $r=\Vert \mathbf{F} \Vert$
\EndWhile
\end{algorithmic}
\caption{Solver using a Newton method for \textcolor{azure}{soft loading}.}
\label{algo:bondbased:solver}
\end{algorithm}

\begin{algorithm}[h!]
\begin{algorithmic}[1]
\State Extend the domain and applied the prescribed displacement $\mathbf{w}$ in the extension.
\State Start with the initial guess of displacement, $\bu_0=0$
\For{ $0\leq k \leq N$ }  
\State Take $\bu^k = \bu^{k-1} + \Delta \bu$
\State Linearize $\mathcal{L}$ such that $\mathcal{L}(\bu^k) = \mathcal{L}(\bu^{k-1}) + H(\bu^{k-1})[\Delta \bu]$
\State Solve for increment $\Delta \bu$ using
\vspace{-0.5cm}
        \begin{equation}\label{eq:pdQuasiNewton:hard}
            H(\bu^{k-1})[\Delta \bu] + \mathcal{L}(\bu^{k-1}+\mathbf{w}) = 0.
        \end{equation}
        \If{ $\Vert \mathbf{w} \Vert - \Vert \bu^k \Vert < \delta$}
        \State Exit loop
        \EndIf
\EndFor
\end{algorithmic}
\caption{Solver using a Newton method for \textcolor{azure}{hard loading}}
\label{algo:newton:hard}
\end{algorithm}

\section{Numerical simulations}
\label{Numerical}
The damage $d(\xx,t)$ at load step $t$ at point $\xx$ in this model is defined as
\begin{align}
    d(\xx,t) = \frac{max(r(t))}{r^c}
\end{align}
where $max(r(t))$ is the largest strain in the neighborhood $H_\epsilon(\xx)$ at load step $t$ and $r^c$ is the strain corresponding to the bond force where bond softening starts. Thus, damage below one $(d < 1)$ indicates that the strain is still in the linear regime. If the damage is equal to one $(d=1)$ the strain has reached the point where softening begin. A damage greater than one $(d>1)$ means that softening started.

\subsection*{Validation against linear elasticity}

\begin{figure}[b]
\centering
\includegraphics[width=\linewidth]{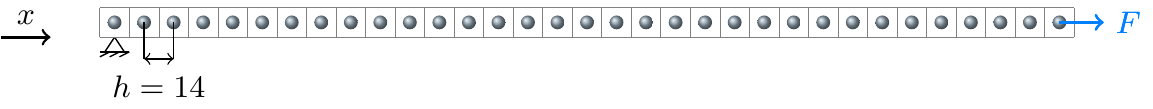}
\caption{Sketch of the one dimensional bar benchmark test. The node on the left-hand side is clamped with respect to displacement $(u=0)$. A force $F$ is applied on the node at the right-hand side. Adapted from~\cite{Diehl2020_validation_1d}.}
\label{fig:sketch:validation:im1}
\end{figure}

For the one-dimensional case, the strain $\boldmath{\epsilon}_\text{CCM}$ from classical continuum mechanics (CCM) is recovered. The stress $\sigma$ is defined as $\sigma=E \cdot \boldmath{\epsilon}$ where $E$ is the material's Young's modulus. The relation of Force $F$ and stress reads as $\sigma=\sfrac{F}{A}$ with $A$ as the area of the cross section. Applying these two relations, the strain is obtained by $\boldmath{\epsilon}=\sfrac{\sigma}{E}=\sfrac{F}{(A\cdot E)}$.  Assuming a force $F$ of \num{40}\si{\newton}, a cross section of \num{1}\si{\square\meter}, and a Young's modulus $E$ of $40$\si{\giga\pascal}, the strain reads as $\boldmath{\epsilon}_{CCM}=\num{1e-8}$.\\

Figure~\ref{fig:sketch:validation:im1} sketches the geometry for the one-dimensional model problem, which is used to recover the strain from classical continuum mechanics. The node on the left-hand side is clamped with respect to displacement. A load in force is applied to the first node. The length of the bar is \num{16}\si{\meter}.  We chose $\delta=3\cdot h$ with $h=1/4$, a length $L=16$ and the tolerance was set to $\delta=\num{1e-11}$. To determine the bond-based material properties 
and since, we are interested in the linear elastic region of the potential, we set $E=\num{40}\si{\giga\pascal}$. 
For more details on the energy equivalence for the one-dimensional bond-based softening model, we refer to~\cite{CMPer-JhaLipton}. For the global strain we get $\boldmath{\epsilon}=\num{1.01e-8}$ which is close to the strain predicted by classical continuum mechanics $\boldmath{\epsilon}_{CCM}=\num{1e-8}$. As a second validation, the same discretized bar was simulated using the Silling's state based model~\cite{silling2007peridynamic} and the assembly of the tangent stiffness matrix using the numerical approximation of derivative as in~\cite{littlewood2016roadmap}. In that case the predicted strain is $\boldmath{\epsilon=\num{1.e-8}}$ using the author's C++ code~\cite{diehl2018implementation}. The python code finished in \num{1.04}\si{\second} (\num{24} iterations) using the presented approach, and the numerical approximation of the tangent stiffness matrix took \num{3.71}\si{\second} (\num{31} iterations).

\subsection*{Soft loading of a pre-notched square plate}
\begin{figure}[tbp]
    \centering
    \includegraphics[width=0.95\linewidth]{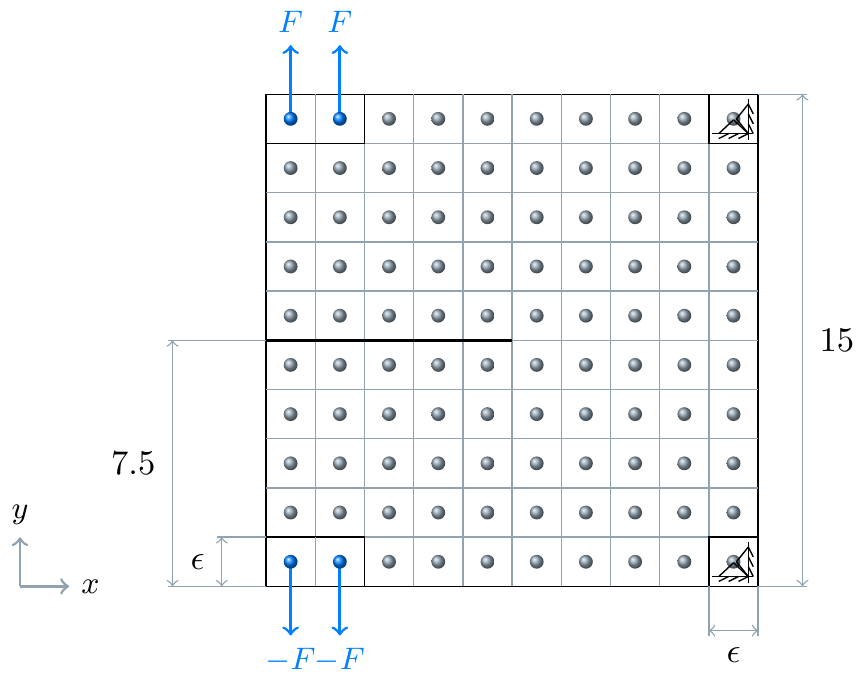}
    \caption{Sketch of the two-dimensional pre-cracked square plate with an initial crack from the mid of the left-hand side to the center of the plate. All nodes in a square of horizon size $\delta$ at the lower right and upper right corner are fixed in displacement in both directions. On the lower left and the upper left, an external force in $y$-direction is applied to all nodes within a rectangle of size $\epsilon\times 13 \epsilon$. Adapted from~\cite{diehl_2020_square}.}
    \label{fig:model:problem:plate}
\end{figure}

Figure~\ref{fig:model:problem:plate} shows a sketch of the pre-cracked square plate (\num{5}$\times\num{5}$) with an initial crack of length \num{7.5}. All nodes within the lower and upper right-hand side square of horizon size $\delta$ are clamped in both directions. All nodes within the lower and upper left-hand side square of horizon size $\epsilon \times 13 \epsilon$ are loaded with the force of $\pm$\num{4e6}\si{\newton} in $y$-direction. The load in force refers to the so-called soft loading in the theory of fracture mechanics. The nodal spacing $h$ was $0.2$ and the horizon $\epsilon$ was $4\cdot h=0.8$. The tolerance was set to $\delta=\num{1e-5}$. All bonds between the PD nodes crossing the initial crack line were removed. As material properties, we chose $E=$\num{30}GPa. 

\begin{figure}[tb]
    \centering
    \begin{subfigure}{0.5\textwidth}
    \centering
    \includegraphics[width=\textwidth]{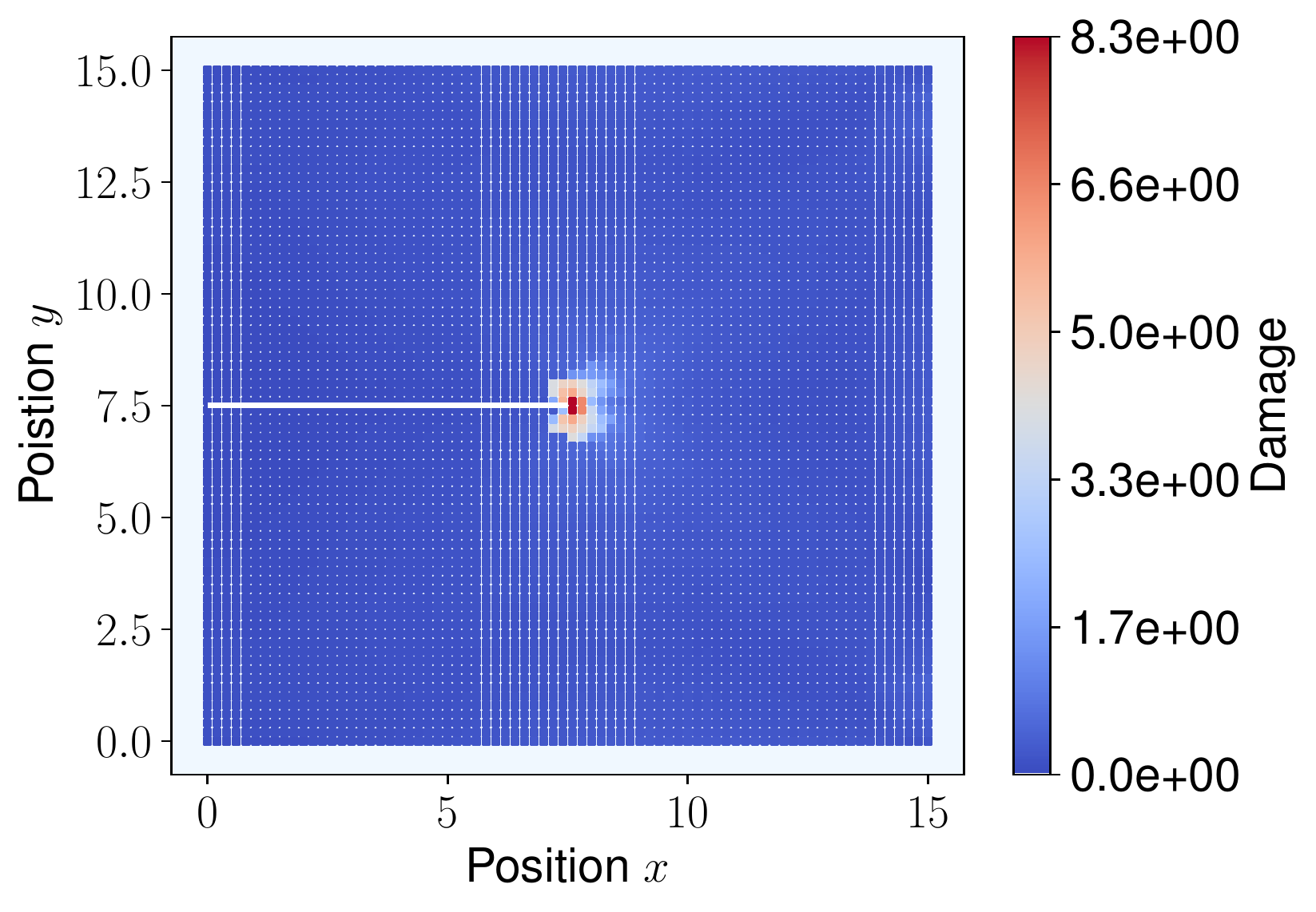}
    \caption{}
    \label{fig:plate:d}
    \end{subfigure}
    
    \begin{subfigure}{0.5\textwidth}
    \centering
    \includegraphics[width=\textwidth]{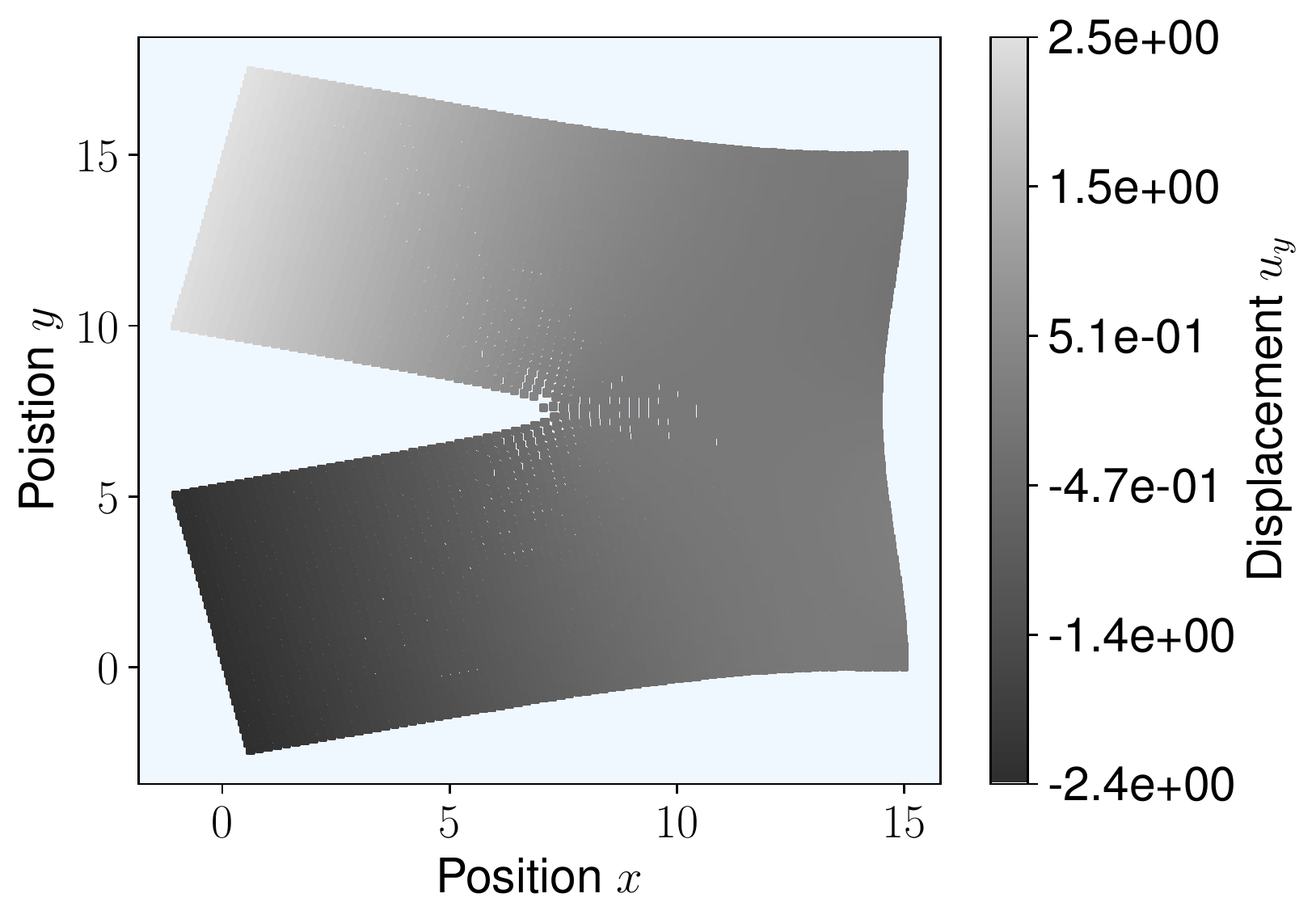}
    \caption{}
    \label{fig:plate:u}
    \end{subfigure}
    
    \caption{Soft loading: Damage (\subref{fig:plate:d}) at the discrete PD nodes (blue = undamaged and red = damaged) the pre-crack is the white line. The displacement (\subref{fig:plate:u}) at the last stable load step.}
    \label{fig:numerical:results:damage}
\end{figure}

A simulation with the following load steps were executed: A external force $F_1=\pm$\num{4e6}\si{\newton} was applied for up to eleven load steps. An external force  $F_2=\pm$\num{4e5}\si{\newton} was applied for one load step after the previous eleven load steps. After that, an external force $F_3=\pm$\num{4e3}\si{\newton} was applied for nine load steps. Note that we had to reduce the load step since the tangent stiffness matrix became unstable with the larger load step. Figure~\ref{fig:plate:d} shows at the last stable load step that damage occurs still only in the neighborhood of the crack tip. 
Figure~\ref{fig:plate:u} shows that the associated deformation remains elastic outside the pre-crack even just before the onset of instability. 

\subsection*{Hard loading of a pre-notched square plate}

\begin{figure}[tb]
    \centering
    \begin{subfigure}{0.5\textwidth}
    \centering
    \includegraphics[width=\textwidth]{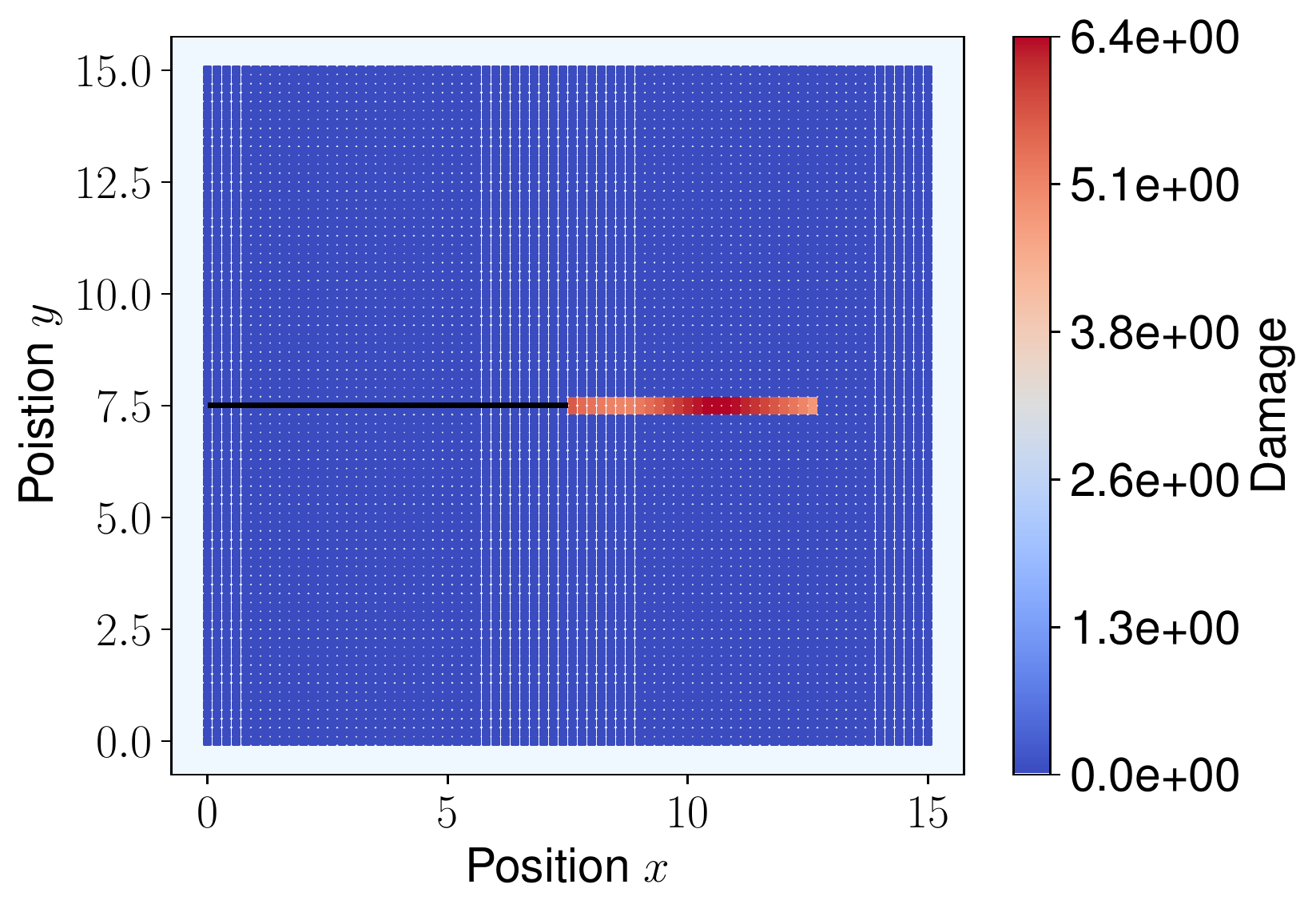}
    \caption{}
    \label{fig:plate:hard:d}
    \end{subfigure}
    
    
    \caption{Hard loading: Damage localizes and concentrates to form a crack. Here, black indicates the pre-crack, (blue = undamaged and red = damaged). The crack is given by the red colored zone is the stable propagating crack from left to right. All other surrounding points are deforming linear elastically.}
    \label{fig:numerical:results:damage:hard}
\end{figure}

For hard loading, we extended the geometry in Figure~\ref{fig:model:problem:plate} with a layer of horizon size $\epsilon$ in $y$-direction. We apply displacement boundary conditions using the prescribed displacement $\mathbf{w}$ and no body force $\bb$ is applied. All simulation details, except the the horizon $\epsilon=8*h$, are kept the same. Figure~\ref{fig:numerical:results:damage:hard} shows the damage field after twelve constitutive load steps. Damage localizes and concentrates to form a crack. Here, black indicates the pre-crack, (blue = undamaged and red = damaged). The crack given by the red colored zone is propagating stably from left to right. All other surrounding points are deforming linear elastically

\section{Conclusion}
\label{Conclusion}
This paper briefly summarizes the ongoing research on peridynamic quasi-static fracture modeling in theoretical and computational directions. For computation, we provide an analytical description of the tangent stiffness matrix and provide the theoretical framework that shows that the solution to the quasi-static equation exists about a stable equilibrium of the peridynamic potential energy for hard and soft loading. Our method for showing this uses fixed point arguments. Examples of the quasi-static crack evolution using the new algorithm are illustrated for both soft and hard loading through numerical examples. 

\bibliographystyle{asmems4}

\section*{Acknowledgments}
PD thanks the LSU Center for Computaiton \& Technology for supporting this work. This material is partially based upon work supported by the U. S. Army Research Laboratory and the U. S. Army Research Office under Contract/Grant Number W911NF1610456.

%
\bibliography{asme2e}

\end{document}